\documentclass{article}

\usepackage[dblblindworkshop, final]{neurips_2026}
\makeatletter
\renewcommand{\@noticestring}{}
\makeatother

\usepackage[utf8]{inputenc} 
\usepackage[T1]{fontenc}    
\usepackage{hyperref}       
\usepackage{url}            
\usepackage{booktabs}       
\usepackage{amsfonts}       
\usepackage{nicefrac}       
\usepackage{microtype}      
\usepackage{xcolor}         
\usepackage{graphicx}

\title{Measuring Human-AI and AI-AI Teams in \\ Multi-turn Layout Creation}
\workshoptitle{Human-AI Coevolution: Measuring Human-Agent Teams in the Agentic Era}

\author{%
  Virdi Rajvir Singh Naginder Singh\\
  College of Computing and Data Science\\
  Nanyang Technological University\\
  \texttt{ravirsi001@e.ntu.edu.sg} \\
  \And
  Yewen Pu \\
  College of Computing and Data Science \\
  Nanyang Technological University \\
  \texttt{yewen.pu@ntu.edu.sg}\\
}

\begin{document}

\maketitle

\begin{abstract}
Co-creating layouts such as websites and slides is a useful testbed for evaluating multi-turn instruction following of LLM systems. 
Since an intended layout's appearance cannot be well expressed in a single turn, the user must wait for the LLM to generate a layout then iteratively refine it with the LLM over multiple turns. 
We present a small-scale, exploratory (n=6) study of this multi-turn process to compare the differences of human-AI and AI-AI teams on the same layout-generation task.
Given a reference wire-frame layout, a participant or a language model `speaker` issues natural-language instructions to a fixed `listener` LLM over multiple turns until the speaker is satisfied.
Rather than an LLM as the judge, we fit a scoring function over geometric layout features using curated human preference data (n=19).
We find that LLMs tend to encode the entire layout in the first turn, while humans iterate over many turns on average, writing much shorter instructions. 
We also find that human-AI teams consistently scored lower than AI-AI teams. 
This gap is driven by a limiter of human participants in specifying precise sizes and positions (e.g. 42 pixels to the right).
Humans perform almost at par with language models on other features like ordering the elements into nested structures.
We then perform interventions on the speaker LLMs to make them more human-like, and find it difficult.
This raises the challenge of using LLMs as user models to evaluate multi-turn co-creation tasks.
\end{abstract}

\section{Introduction}

\begin{figure}
  \includegraphics[width=1.0\textwidth]{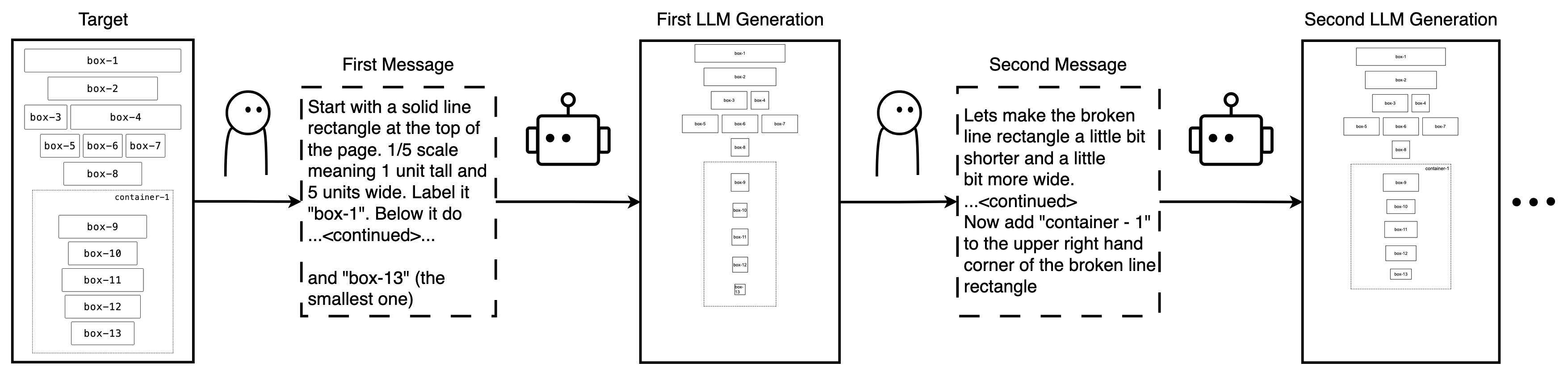}
  \caption{Example of first two turns of an interaction between a participant and the LLM. Individual messages may be purely creation of new elements (Message 1) or purely adjustment of existing elements (Message 2)}
  \label{fig:task}
\end{figure}

Increasingly, designers are generating layout artifacts such as web pages and flowcharts using LLMs.
Since it is difficult to precisely specify a layout using language upfront, a human typically describes a rough gist of the layout, gets an initial response from the LLM, then refine it over multiple turns toward the desired output.
Consequently, layout generation provides a natural evaluation for LLM's ability on multi-turn instruction following.
In these tasks, a human participant is given a visual target layout, then converses with an LLM agent over multiple turns to recreate it (Figure \ref{fig:task}).
Performance is measured as the similarity between the LLM's output and the target design.

Some prior works on multi-turn layout generations have either measured performance using another large model (LLM-as-judge) \citep{si2024design2code} or similarity using neural encoders (e.g. Clip embedding) \citep{radford2021clip} \citep{si2024design2code}.
Similarities measured this way have not been verified to match how humans would judge how two layouts would be similar.
Other works have used LLMs as proxy users rather than human participants \citep{zhou2024webarena}, which miss the nuances that come from naturally given language instructions over multiple turns.

This paper presents a small-scale study of human-AI teams in layout generation, evaluated using a similarity metric fitted to real human judgments, and a comparison between AI-AI teams performing the same task. 
We first constructed 5 simple wire-frame like layouts that were presented to 6 participants. The instructions, raw generations, and session information including time spent was collected to populate the corpus which was then rated using the metric derived using a second Binary Preference Elicitation study consisting of 19 participants and 55 generation pairs. 
We find that while humans tend to construct the design over multiple turns, LLMs tend to commit an entire layout in a single turn.
We also find that humans are more efficient than the flagship speaker models, typing roughly a third as many characters per instruction. 
However, this efficiency does not translate to better end results --- LLM speakers achieve higher overall scores when compared to humans.
Analyzing the human-AI trajectories, we find humans are poor at specifying exact sizes and positions using natural language, yet communicate general spatial relationships well.

This work makes the following contributions:
\begin{itemize}
    \item A small-scale corpus of real, iterative human layout-generation sessions paired with a matched language-model-speaker.
    \item An interpretable similarity metric fit from real human pairwise preference judgments over geometric and structural features on contentless layouts.
    \item An empirical account of where human layout generations fall short of language models: in the precise spatial matching of individual elements.
\end{itemize}

\section{Related Work}
\label{sec:related}

\paragraph{Human-human design communication.} Closest to our human-side
design is McCarthy et al.'s study pairing novice participants as Designer
and Maker to recreate target CAD designs over multiple rounds using
drawing and text \citep{mcCarthy}, and its large-scale
follow-up mrCAD, a dataset of 6082 such communication games across
1092 pairs that explicitly separates generation instructions from
refinement instructions \citep{mccarthy2025mrcadmultimodalrefinementcomputeraided}. Both studies also
compare a model's interpretation of the same instructions against
additional human participants performing the identical role (GPT-4V as a
``Solo Maker''). These works focus on a richer multi-modal input, while we focus on natural language instructions which is a common medium in human-AI teams. 

\paragraph{Generating designs from a visual target.} Pix2Code
\citep{beltramelli2017pix2code} first mapped UI screenshots to code;
WebSight \citep{laurenccon2024websight} trains image-to-HTML models on
synthetic pages, and Design2Code \citep{si2024design2code} benchmarks
screenshot-to-code conversion with multimodal models. These establish the
task but treat it as single-turn: a target is given, output is produced
once, with no interactive refinement. Where multi-turn is studied, it is typically a single LLM agent acting directly, taking its own actions to complete a
long-horizon task, rather than instructing a separate system over
multiple turns \citep{zhou2024webarena}. Design2Code is the closest prior work to our
evaluation: it also collects binary pairwise human judgments of which of
two candidates is closer to a reference, and fits a logistic regression
over metric differences to predict them, reaching 76.9\% held-out accuracy.
We adopt the same Bradley-Terry formulation \citep{bradleyterry1952}, which is
logistic regression on feature differences, but over a disjoint feature set. Design2Code's dimensions are dominated
by content and appearance (text similarity, color difference, CLIP
embedding), which are undefined on our contentless wire-frame stimuli by
construction. We instead fit purely geometric and structural features,
isolating what raters weigh when content and appearance are held constant. Unlike prior multi-turn work, we also run a matched language model as a speaker setting to analyze human-vs-ai speakers.
\section{Method}

\subsection{Stimuli}
\begin{figure}
  \includegraphics[width=1.0\textwidth]{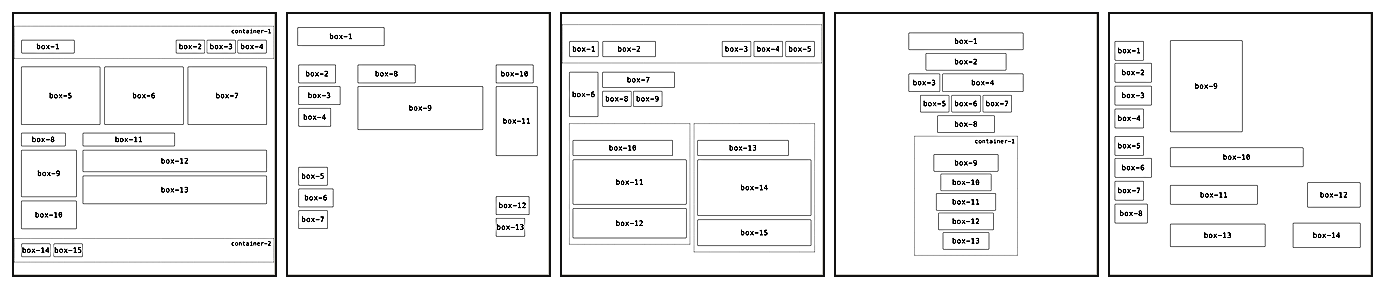}
  \caption{All reference stimuli used in the study. Common features of the structure includes nested structures and relative positions and sizes}
  \label{fig:stimuli}
\end{figure}

The Stimuli/references as seen in figure \ref{fig:stimuli} consist of 5 simple HTML web pages inspired by CSSZenGarden \citep{shea2003csszengarden}. These web pages have no content and color since the focus of the study is on structures like nested elements, relative positioning and sizing. They contain rectangular elements labeled in the form box-\textit{i} where \textit{i} refers to the index number. Dotted border elements container-\textit{i} represent headers, sidebars and nests found in real web pages. These containers house one or many boxes inside them. 

\subsection{Task}
\label{sec:task}
The task objective is to interact with a vision language model (Claude Opus 4.8) across multiple back and forth turns to refine the generated web pages and get it as close as possible to the target web page. At each step they must look at the generated output by the language model, compare it with the target, and decide to make adjustments, introduce new elements, or choose to end the task if they are satisfied by the closeness to the target. No restrictions were placed on the strategy, language or word count of individual instructions.

\subsection{Collection}
\label{sec:h2acollection}
Each participant entered two 15-minute sessions (each allowing a one-time 5-minute extension) where they are shown two individual random references as the targets.

\paragraph{Collect platform design} We collected full instructions, generated HTML and session start and end times for further analysis. The vision language model setup was kept minimal to mirror real coding agents. To ensure consistent generation and formatting of the page for downstream feature extraction at \ref{sec:features}, the model's system prompt was tuned to output black\&white HTML following the box-\textit{i} container-\textit{i} labeling conventions. 

\subsection{Metric}
We define a single similarity score for a generated page against its reference. Boxes and containers in the generated page are matched to their counterparts in the reference by their shared label conventions. Over this matched set we computed a small set of interpretable geometric features (\S\ref{sec:features}) represented by the vector \(\phi\). The metric is a weighted sum of these features, where the weights are derived from the binary preference elicitation sub study \ref{sec:pe}. The final score is normalized within [0,1] using the theoretical maximum obtained by feeding the same layout as both the reference and the generation (maximum closeness as the elements are identical).

\subsubsection{Features}
\label{sec:features}
The feature set starts from 14 candidate geometric features computed over the matched box-container elements. Upon computing the individual feature scores for all generations collected in \ref{sec:h2acollection}, the following elements were removed because of zero-variance across the entire corpus:
\begin{itemize}
    \item f2 (precision) - measures the fraction of generation's labeled elements that also exist in the reference. Removed because participants never generated non-existent elements and strictly followed the box-container convention and indices.
    \item f11 (alignment delta) - measures consistency of elements aligning to shared edges.
    \item f12 (overlap delta) - a whole-layout measure of pairwise element overlap.
\end{itemize}

The following set of features produced a high correlation with other features and were excluded due to a lower computed weight (at \ref{sec:weights}) compared to their correlated feature:
\begin{itemize}
    \item f3 (area recall) - high correlation at \(r=0.85\) with f1.
    \item f4 (euclidean position error) - high correlated at \(r=0.78\) with f9.
    \item f6 (relative width match) - high correlation with f8 (aspect-ratio match).
    \item f10 (quadrant agreement) - correlation with f4, f1 and f3.
    \item f13 (group count consistency) - high correlation with f1.
\end{itemize}

This step leaves the following feature set \(\phi\):
\begin{itemize}
    \item \textbf{f1 (recall)} - fraction of the reference's labeled elements that were reproduced at all.
    \item \textbf{f7 (relative height match)} - how closely a matched element's height relative to page height matches the element.
    \item \textbf{f8 (aspect-ratio match)} - how closely a matched element's height-to-width ratio matches the reference's, averaged over all matched elements.
    \item \textbf{f9 (reading-order consistency)} - for every pair of matched elements, whether their left-right and top-bottom ordering agrees between reference and generation, averaged over both axes and all pairs.
    \item \textbf{f14 (container uniformity)} - whether recall is even across the reference's container groupings, rather than concentrated in some containers and absent from others. 
\end{itemize}

Similarly f5 (Chebyshev position error) was removed earlier because of a near duplicate computation to f4.

\subsubsection{Preference elicitation task}
\label{sec:pe}

A Bradley-Terry model was used to fit the weights above using human pairwise preference judgments over real generations. For each of the 5 references, 5 real candidate generations are evenly sampled from actual participant sessions. Candidates are paired exhaustively within each reference (10 real pairs per reference, all \(C(5,2)\) combinations), plus one gold pair per reference. The gold pair was specifically structured as a strong real candidate paired with an empty "broken" candidate. This was used as an attention check to prevent participants from choosing within a pair at random.

A rater is shown the reference layout alongside two candidate renderings and makes a single binary choice: which candidate is closer to the reference.

\subsection{Language model speakers}
\label{sec:a2a}
To compare how human-AI and AI-AI teams differ on this task, we run a second condition in which the speaker to the vision language model is replaced by a second vision language model. The instructions to this new speaker were kept identical. Any differences in the interactions and outcomes between this condition and the human study are attributable to how the speaker communicates.

Three driver models take the speaker role, chosen to span a range of capability and cost: Claude Opus 5 (frontier model), Gemini 3 Flash Preview (mid-tier), and Qwen2.5-VL-72B-Instruct (cheap, open-weight). 

\section{Result}

\subsection{The corpus}

\paragraph{Collected data} The layout-generation study produced around 90 total turns/instructions with up to a maximum of 17 turns in a single session. Of 18 respondents that participated, 8 completed both generation sessions; 11 sessions by 6 participants met the inclusion criteria of having at least 2 matched boxes. Each reference was targeted by two separate participants to limit individual outliers dominating it. The preference elicitation study consisting of 19 participants taking on 55 judgments each with a range of 2-8 minutes per participant and a per-judgment response time of about 4.5 seconds. A total of 718 decisive pairwise judgments which excludes ties and gold attention-check pairs was produced by this study. The mean pairwise rater agreement on the 50 real pairs was 0.891. The full corpus is available at \url{https://virdirajvir.github.io/RLG/}.

\begin{figure}[!h]
  \includegraphics[width=1.0\textwidth]{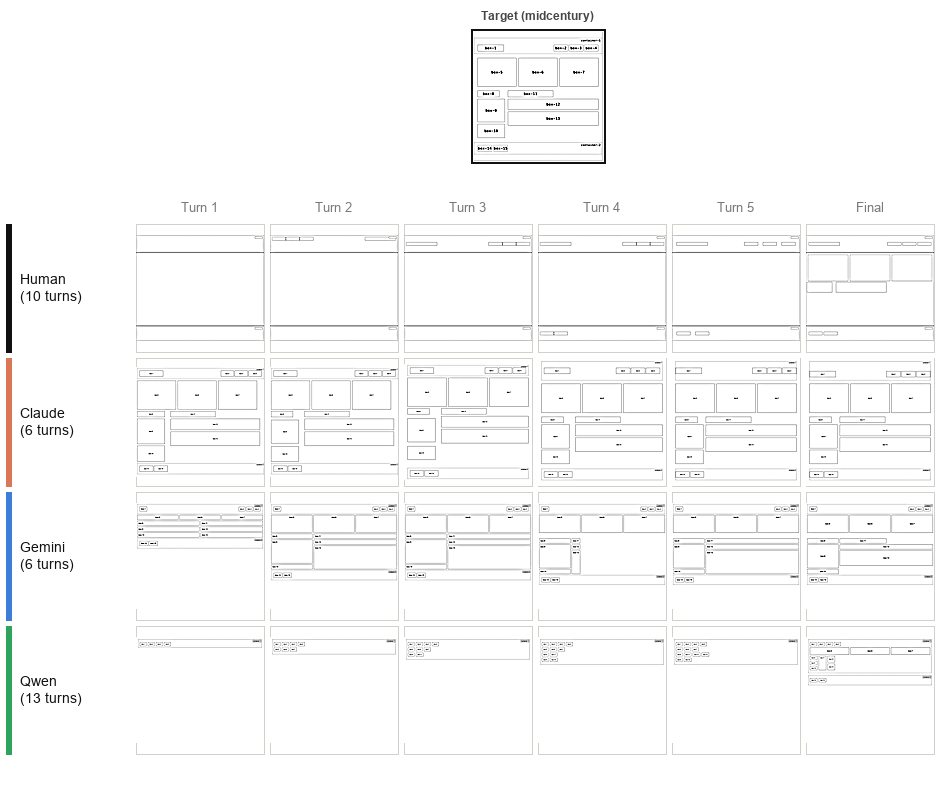}
  \caption{Turn by Turn progress difference between speakers on the same reference. Claude and Gemini exhibit front-loading: creating all elements in the first turn}
  \label{fig:fig2}
\end{figure}

\paragraph{Example session trajectories} Figure \ref{fig:fig2} shows one reference tracked turn by turn across all four speakers (human, Claude, Gemini, Qwen) in one session on a fixed target reference named midcentury. 

\subsection{Weight Fitting}
\label{sec:weights}
We identified 7 feature pairs with an absolute correlation of 0.7 or higher on the training split (80\%, or 574 judgements) and for each pair removed the member with the smaller fitted coefficient magnitude. The trimmed five-feature model is both more accurate (\(87.9\% \) vs \(87.3\%\) mean accuracy across the folds) and more stable (\(std \space 0.8\) vs \(std \space 1.5\) points) across these folds than the ten-feature model, and will be used in the analysis. 

The final weights are fit on the complete set of 718 judgments for the most precise coefficient estimates:
\begin{itemize}
    \item f1 (recall) - \(+4.59\),
    \item f8 (aspect-ratio match) - \(+1.62\), 
    \item f9 (reading-order consistency) - \(+0.70\),
    \item f14 (container uniformity) - \(+0.40\),
    \item f7 (relative height match) - \(0.002\), not significant with 95\% CI [-0.25, 0.24], comfortably spanning zero weight, but kept for consistency with the feature set validated by the trimming procedures above.
\end{itemize}

Leave-one-rater-out cross-validation on this final fit gives 88.3\% accuracy, against a 75.8\% baseline of simply picking the next generation turn.  

\subsection{Similarity score over turns} 
\begin{figure}[!h]
  \includegraphics[width=1.0\textwidth]{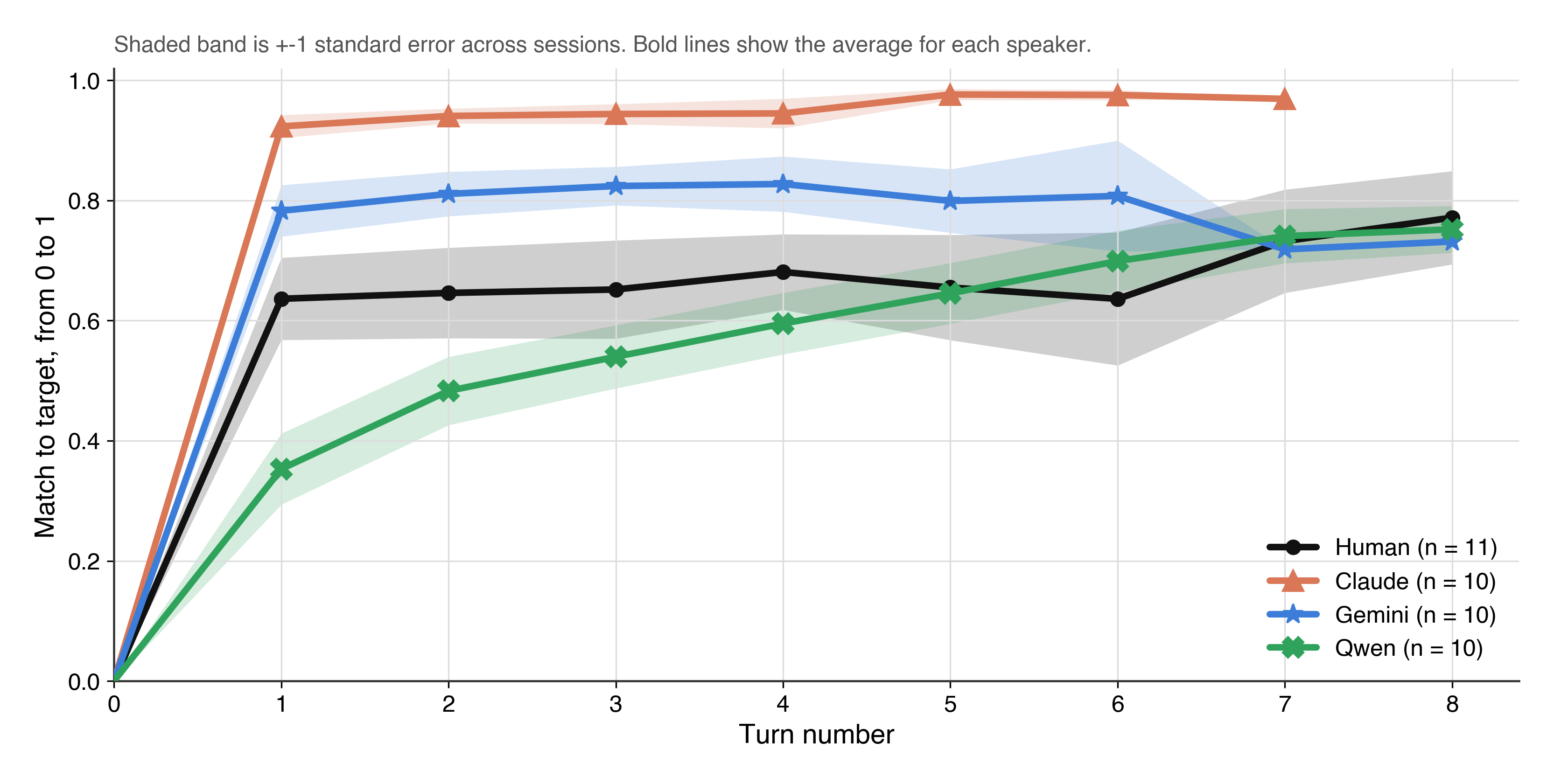}
  \caption{Closeness-to-target score trends vs turn across Human, Claude, Gemini and Qwen. The shaded tunnels represent standard error.}
  \label{fig:fig4}
\end{figure}

Using the weights derived at \ref{sec:weights}, figure \ref{fig:fig4} quantifies the closeness score pattern across turns, truncated to the first 8 turns. 

\subsection{Analysis}
\label{sec:analysis}

\begin{figure}[!h]
  \includegraphics[width=1.0\textwidth]{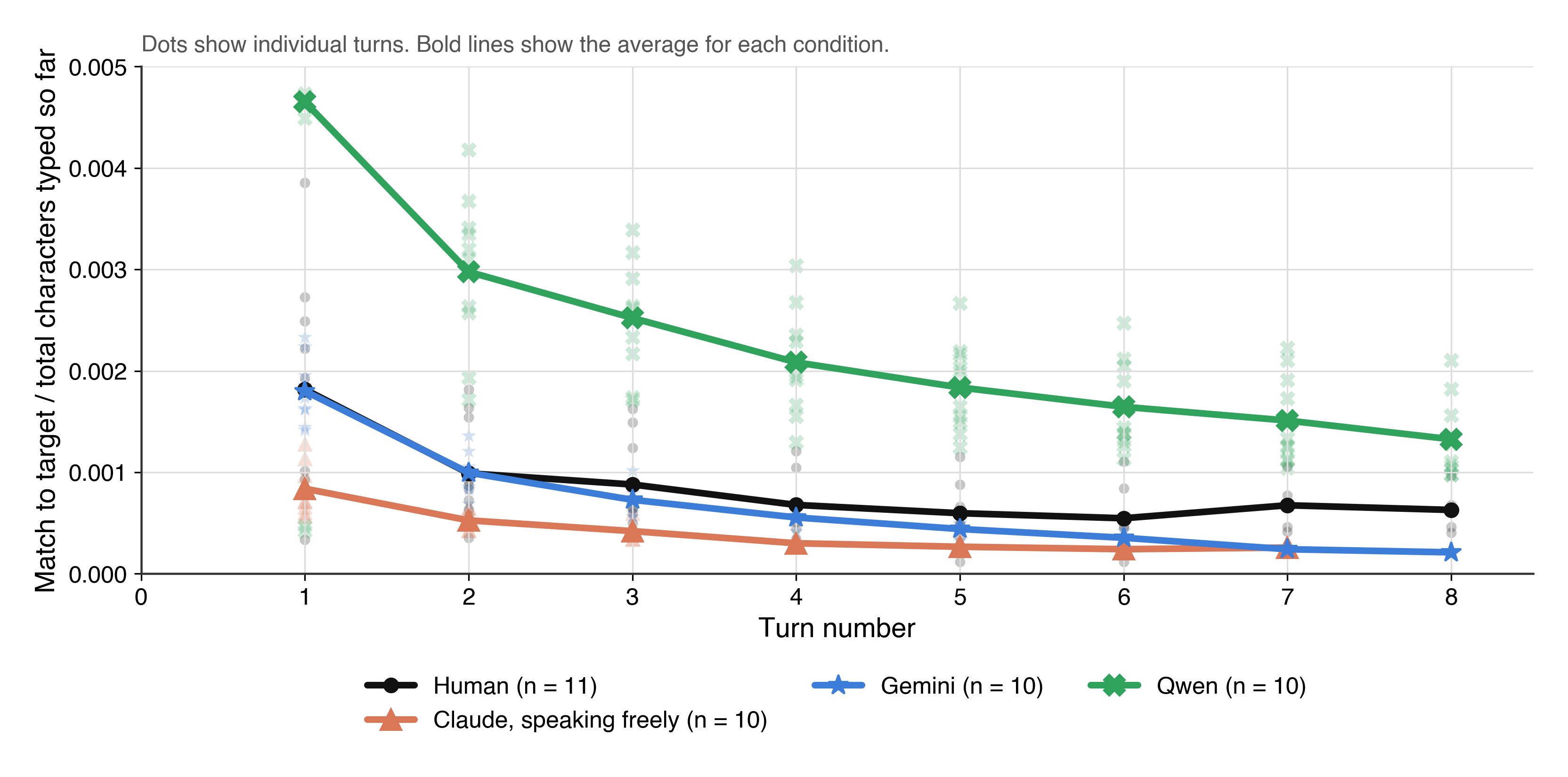}
  \caption{Closeness-to-target score per total characters typed so far. Claude trajectories represent dense instructions while Humans, Gemini and especially Qwen uses fewer characters over longer trajectories}
  \label{fig:efficiency}
\end{figure}

Figure \ref{fig:efficiency} plots match-to-target score against cumulative characters typed, to see how much cumulative progress each speaker buys on an average per character of instruction, split over turns. Claude's curve sits well below the others on this axis: averaging over 700 characters per instruction, while humans (214), Gemini (407) or Qwen (84) sit much higher. Though Claude sessions are shorter compared to Qwen sessions, the total characters written by Claude is still greater compared to Qwen, highlighting a best score attainable vs efficient cost usage tradeoff. 

\begin{figure}[!h]
  \includegraphics[width=1.0\textwidth]{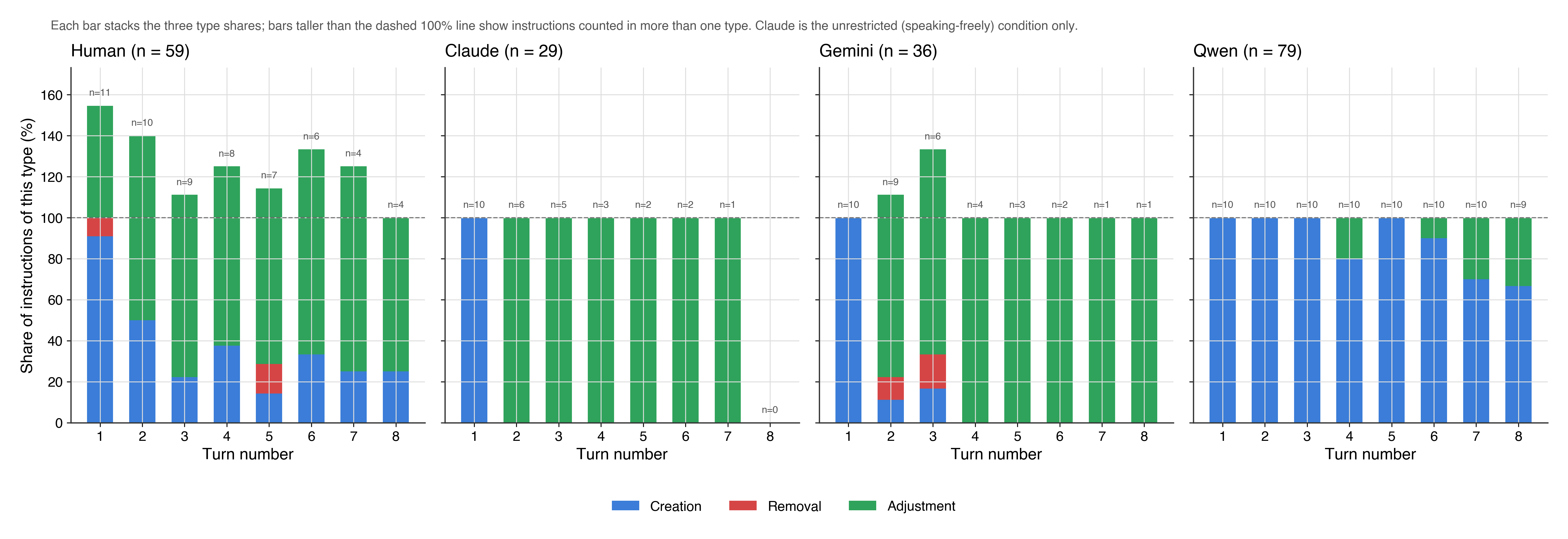}
  \caption{Percentage share of instruction intent types per turn. Claude and Gemini exhibit front-loading while humans and Qwen take an incremental approach}
  \label{fig:category-share}
\end{figure}

\begin{figure}[!h]
  \includegraphics[width=1.0\textwidth]{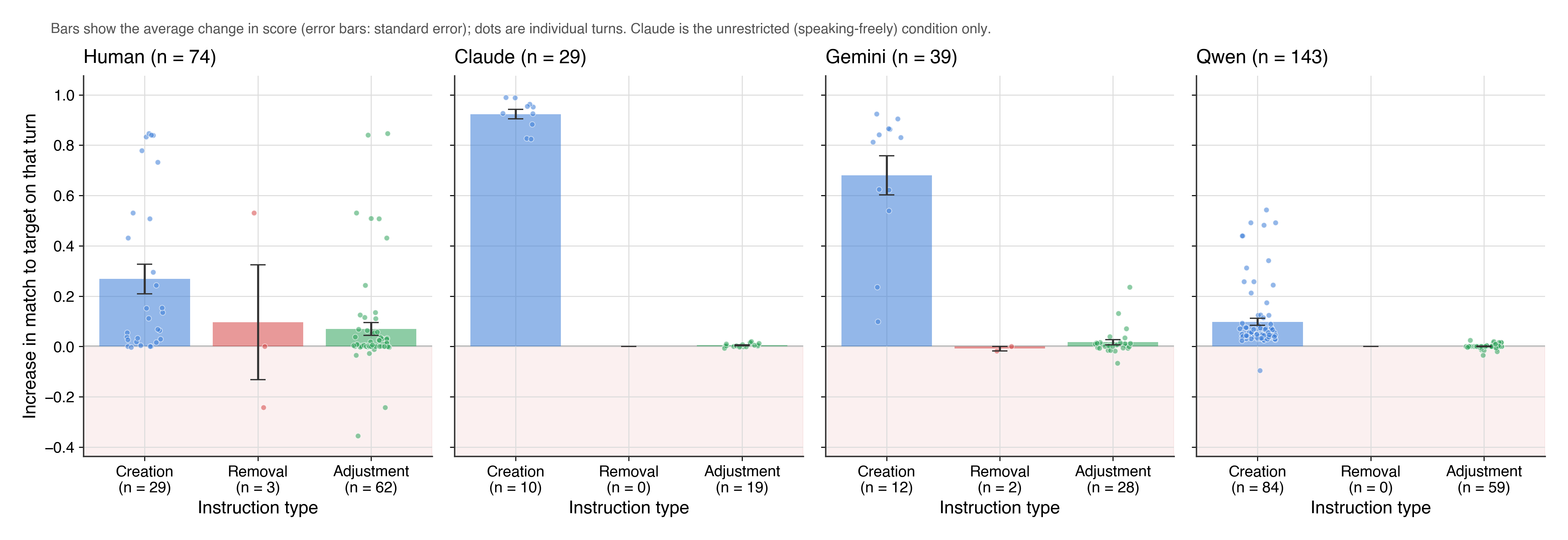}
  \caption{Change in score grouped by instruction intent. Humans choose a strategy of generic creation and later adjustments}
  \label{fig:category-gain}
\end{figure}

Figures \ref{fig:category-share} and \ref{fig:category-gain} break this difference down by instruction intent. Instruction intent was classified into Creation, Removal, and Adjustment using a simple Claude Haiku classification on all turns and having an 87\% classification accuracy on manually self-labeled turns. Claude's first turn is 100\% creation and nothing else while every turn after the first is 100\% adjustment. Gemini is nearly identical while Qwen is the opposite. Human sessions sit in between: turn 1 is mostly but not exclusively creation (91\%, with more than half of turn-1 instructions also containing an adjustment in the same message), creation share then drops and fluctuates turn to turn while adjustment stays consistently high (86-100\%) for the rest of the session. The percentages going over 100\% are due to non-exclusive classification: some instructions create and adjust at the same time for example.

Figure \ref{fig:category-gain} shows what this buys each speaker. For Claude and Gemini, creation-only turns account for 99.0\% and 94.3\% of total score gain. Qwen's creation turns account for 99.5\% of its gain but each one buys far less (mean gain of only 0.10 each). Humans are the outlier: creation-only turns account for 47.5\% of total gain and mixed creation-plus-adjustment turns account for another 40.4\%, while pure adjustment turns, the single most frequent turn type by count (43 of 74 classified human turns), contribute only 8.5\% of total gain, at a mean of 0.016 score points per turn.

Figure (Appendix, \ref{fig:elements-per-instruction}) shows how Claude and Gemini reference an average of 9.1 and 9.9 elements per instruction respectively while human and Qwen instructions reference only 3.3 and 3.2 elements on average.

\begin{figure}[!h]
  \includegraphics[width=1.0\textwidth]{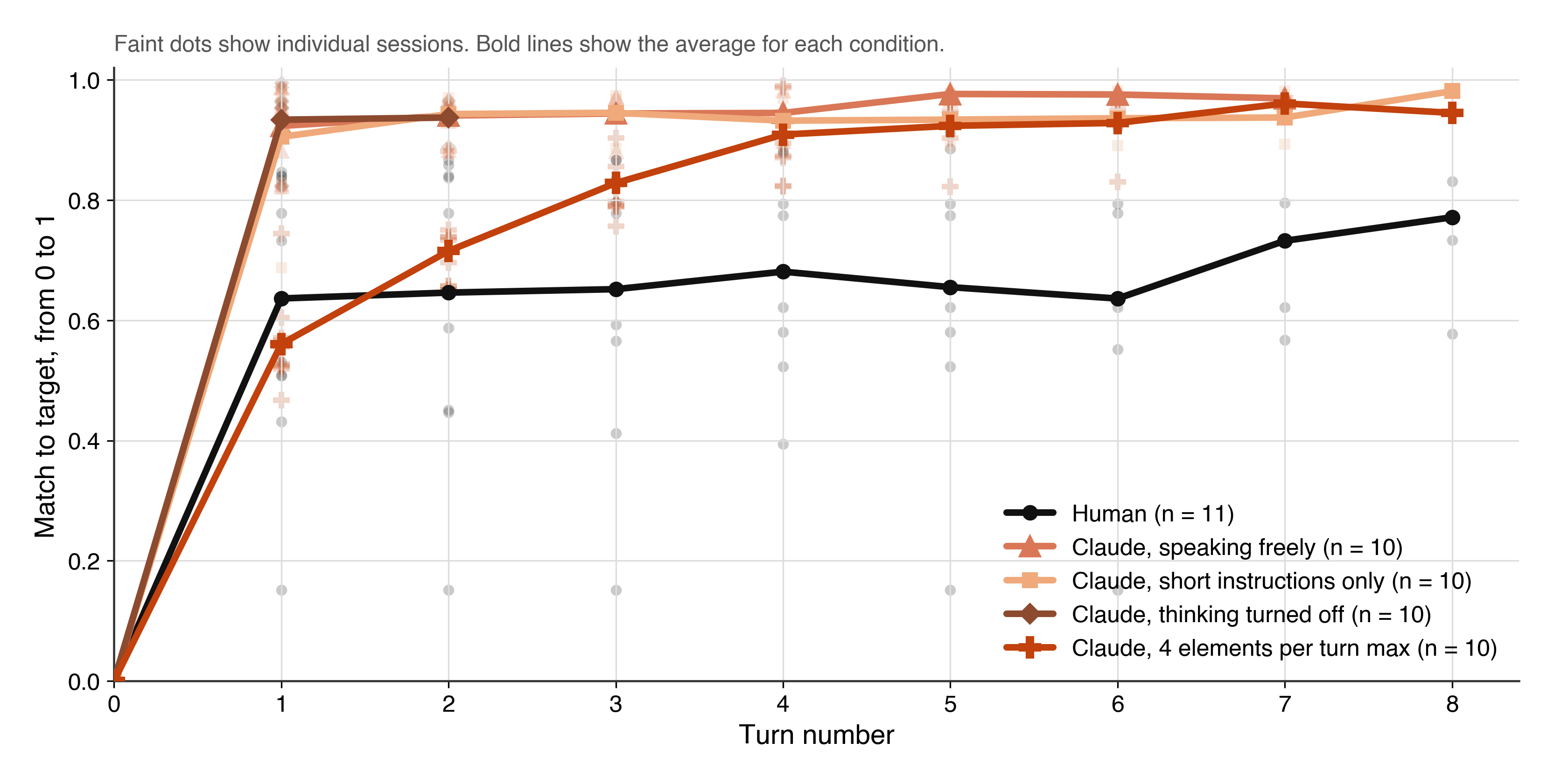}
  \caption{Closeness-to-target score for the four Claude conditions and humans. Limiting Claude to reference only 4 elements per instruction changes its trajectory to a gradual slope}
  \label{fig:claude-ablations}
\end{figure}

Claude was restricted on its word count, its thinking capabilities and on the number of elements it can refer to in a single turn to measure where its advantage is derived from. Figure \ref{fig:claude-ablations} tests these conditions against human. Capping instruction length or disabling extended thinking barely changes anything. The word cap was fully complied with (all 32 capped instructions fall under 200 words) but only affects 17.2\% of uncapped turns. Capping the number of elements Claude may reference per instruction to four is different: turn-1 completion falls to 60\% of the eventual score, and the average session lengthens to 5.4 turns with a gradual increment. What doesn't change is where these four conditions end up: terminal scores are all around 0.930 (vs human's 0.749 terminal score).

Another condition subjected to Claude was to apply few-shot learning to Claude's system prompt based on a random sample of instructions from all 6 human participants. However, this test produced no significant shift in the slope or terminal scores (Appendix, \ref{fig:fewshot}).

\begin{figure}[!h]
    \centering
    \includegraphics[width=1.0\linewidth]{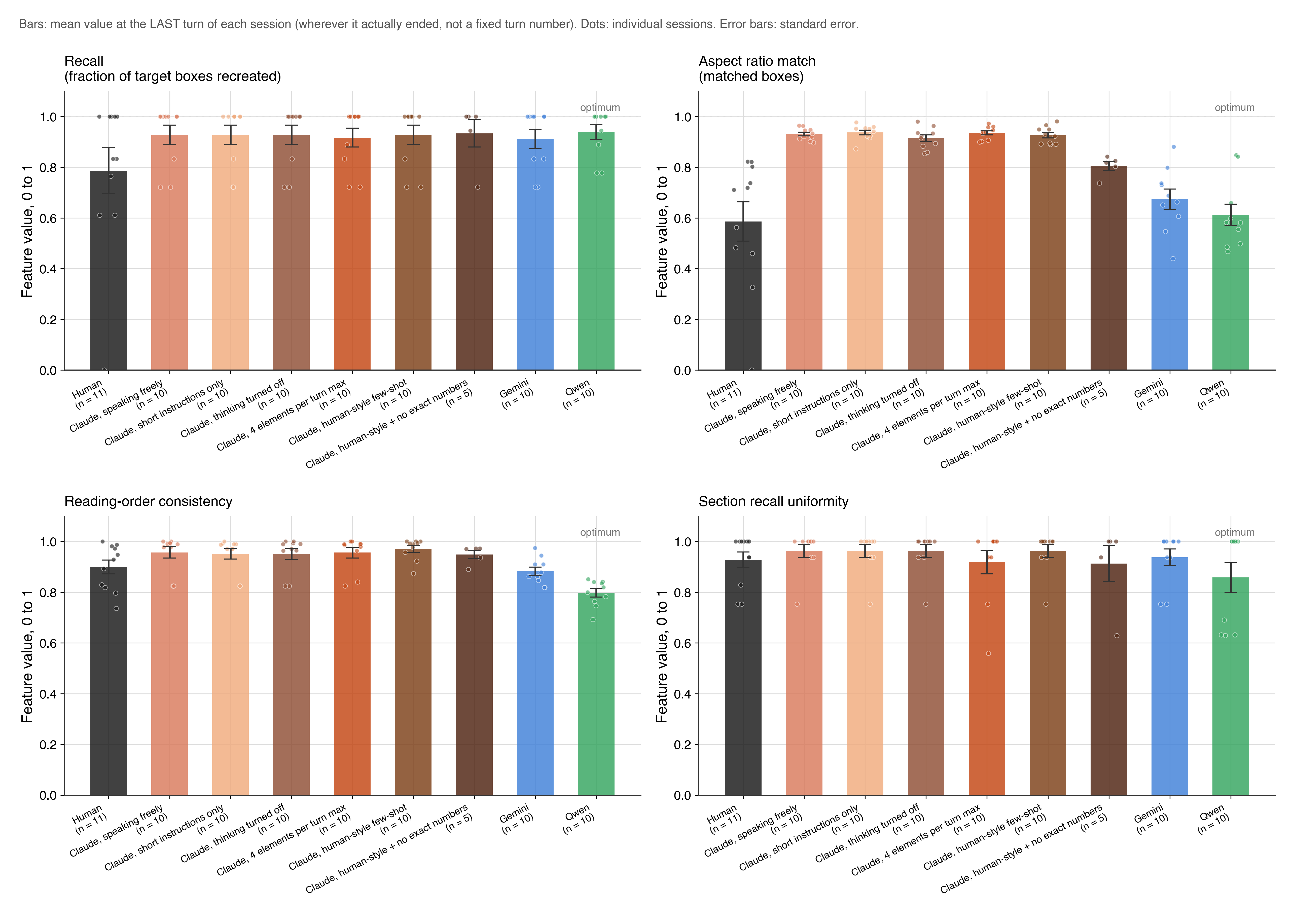}
    \caption{Feature breakdown of terminal scores divided by all conditions and humans. Aspect-ratio matching shows where the structural limiter for humans lies.}
    \label{fig:feat-breakdown}
\end{figure}

\begin{figure}[!h]
  \includegraphics[width=1.0\textwidth]{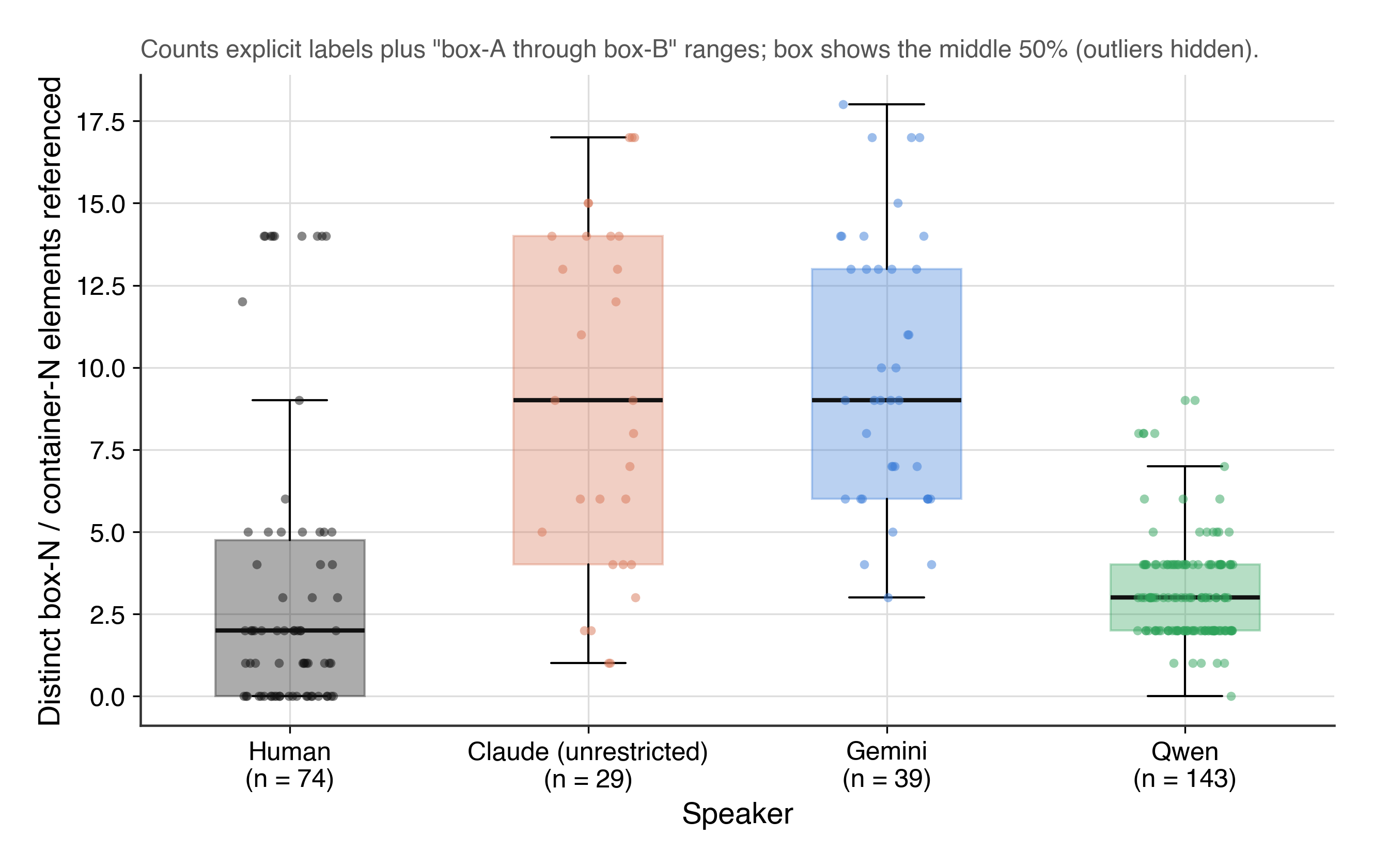}
  \caption{Box whisker plot demonstrating how many target elements models and humans reference on an average per turn. Claude and Gemini reference many more elements per turn, indicating their detailed prompts with many creations and adjustments at once. Humans and Qwen on the other hand focus on lesser elements per turn}
  \label{fig:elements-per-instruction}
\end{figure}

\begin{figure}
    \centering
    \includegraphics[width=1.0\linewidth]{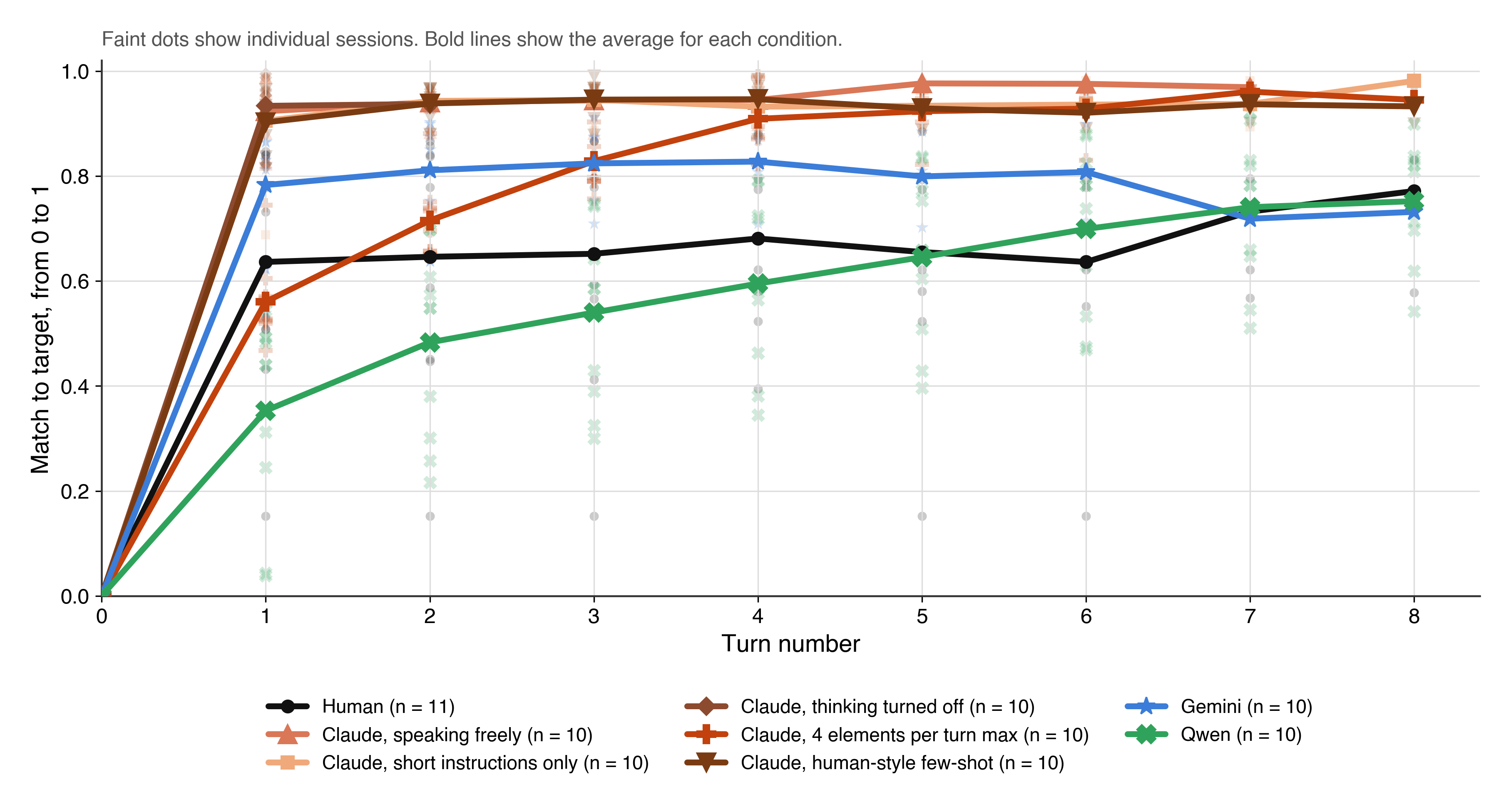}
    \caption{Few-shot learning Claude along with other conditions. No significant deviation from unrestricted Claude trajectories}
    \label{fig:fewshot}
\end{figure}

The terminal scores for each of the 4 significant features (f1 - recall, f8 - aspect-ratio match, f9 - reading-order consistency and f14 - container uniformity) across all the language model conditions and humans is shown in figure \ref{fig:feat-breakdown}. Aspect-ratio match is the most significant differential where Claude scores substantially higher than the rest of the conditions and humans, and it is a structurally different feature that focuses on relative sizing of matched elements compared across the reference and the generated layout. 

\section{Conclusion and Future works}
\label{sec:discussion}
The key result is that human-AI teams converge to a lower match-to-target score than AI-AI teams.
The difference in scores is concentrated in the aspect-ratio matching feature, on which human-AI teams score lower than every model condition. This is consistent with natural language being better suited to conveying holistic structure than the finer proportions of individual elements.

Comparing human-AI and AI-AI trajectories reveals that humans communicate layouts incrementally, while LLMs (except Qwen) tend to front-load the entire layout in the first turn. 
We also find it difficult to coax LLMs to generate human-like instructions, raising a challenge of using LLMs as user simulators for layout generation tasks.

To determine if the human trajectory ceiling is unique to how language models interpret qualitative sizing instructions, a similar study could be conducted with human-human teams. 
To determine if the LLM-generated instructions are intuitive for humans to understand, one may also conduct additional AI-human teams, where AI instructs a human to create the layout.

\section{Limitations.} The study is small in absolute terms (6 qualifying participants for analysis against a single fixed
builder model) so these patterns are not yet established at a scale that
supports strong generalization claims. The scoring metric was fit entirely
on human-generated candidates (\S\ref{sec:pe}) and its transfer to AI-generated pages is not directly tested, since we collected no preference judgments over AI output. As a partial check, 83.8\% of the 420 AI-generated turns lie inside the per-feature range spanned by the elicitation candidates, the main exception being aspect-ratio match, where 12.1\% of AI generations exceed the fitted maximum, though by at most 0.050. Because the model is linear in the features, this is mild extrapolation rather than an unconstrained one --- but it falls on the feature carrying our central comparison, so that result in particular should be read as exploratory.

\section{Acknowledgment}
This research / project is supported by the NTU NAP Start Up Grant Award number \#025535-00001 and NTU URECA
Undergraduate Research Programme

\clearpage
\bibliographystyle{plainnat}   
\bibliography{references}      

\begin{thebibliography}{9}
\providecommand{\natexlab}[1]{#1}
\providecommand{\url}[1]{\texttt{#1}}
\expandafter\ifx\csname urlstyle\endcsname\relax
  \providecommand{\doi}[1]{doi: #1}\else
  \providecommand{\doi}{doi: \begingroup \urlstyle{rm}\Url}\fi

\bibitem[Beltramelli(2017)]{beltramelli2017pix2code}
Tony Beltramelli.
\newblock pix2code: Generating code from a graphical user interface screenshot.
\newblock In \emph{Proceedings of the ACM SIGCHI Symposium on Engineering Interactive Computing Systems}, 2017.

\bibitem[Bradley and Terry(1952)]{bradleyterry1952}
Ralph~A. Bradley and Milton~E. Terry.
\newblock Rank analysis of incomplete block designs: I. the method of paired comparisons.
\newblock \emph{Biometrika}, 39\penalty0 (3/4):\penalty0 324--345, 1952.
\newblock URL \url{https://www.jstor.org/stable/2334029}.

\bibitem[Laurençon et~al.(2024)Laurençon, Tronchon, and Sanh]{laurenccon2024websight}
Hugo Laurençon, Léo Tronchon, and Victor Sanh.
\newblock Unlocking the conversion of web screenshots into html code with the websight dataset, 2024.
\newblock URL \url{https://arxiv.org/abs/2403.09029}.

\bibitem[McCarthy et~al.(2024)McCarthy, Matejka, Willis, Fan, and Pu]{mcCarthy}
William~P. McCarthy, Justin Matejka, Karl~D.D. Willis, Judith~E. Fan, and Yewen Pu.
\newblock Communicating design intent using drawing and text.
\newblock In \emph{Proceedings of the 16th Conference on Creativity \& Cognition}, C\&C '24, page 512–519, New York, NY, USA, 2024. Association for Computing Machinery.
\newblock ISBN 9798400704857.
\newblock \doi{10.1145/3635636.3664261}.
\newblock URL \url{https://doi.org/10.1145/3635636.3664261}.

\bibitem[McCarthy et~al.(2025)McCarthy, Vaduguru, Willis, Matejka, Fan, Fried, and Pu]{mccarthy2025mrcadmultimodalrefinementcomputeraided}
William~P. McCarthy, Saujas Vaduguru, Karl D.~D. Willis, Justin Matejka, Judith~E. Fan, Daniel Fried, and Yewen Pu.
\newblock mrcad: Multimodal refinement of computer-aided designs, 2025.
\newblock URL \url{https://arxiv.org/abs/2504.20294}.

\bibitem[Radford et~al.(2021)Radford, Kim, Hallacy, Ramesh, Goh, Agarwal, Sastry, Askell, Mishkin, Clark, Krueger, and Sutskever]{radford2021clip}
Alec Radford, Jong~Wook Kim, Chris Hallacy, Aditya Ramesh, Gabriel Goh, Sandhini Agarwal, Girish Sastry, Amanda Askell, Pamela Mishkin, Jack Clark, Gretchen Krueger, and Ilya Sutskever.
\newblock Learning transferable visual models from natural language supervision.
\newblock In \emph{International Conference on Machine Learning}, pages 8748--8763, 2021.

\bibitem[Shea(2003)]{shea2003csszengarden}
Dave Shea.
\newblock {CSS} {Z}en {G}arden: {T}he {B}eauty of {CSS} {D}esign.
\newblock \url{https://www.csszengarden.com/}, 2003.
\newblock Accessed 2026-10-05.

\bibitem[Si et~al.(2024)Si, Zhang, Yang, Liu, and Yang]{si2024design2code}
Chenglei Si, Yanzhe Zhang, Zhengyuan Yang, Ruibo Liu, and Diyi Yang.
\newblock Design2code: How far are we from automating front-end engineering?, 2024.
\newblock URL \url{https://arxiv.org/abs/2403.03163}.

\bibitem[Zhou et~al.(2024)Zhou, Xu, Zhu, Zhou, Lo, Sridhar, Cheng, Ou, Bisk, Fried, Alon, and Neubig]{zhou2024webarena}
Shuyan Zhou, Frank~F. Xu, Hao Zhu, Xuhui Zhou, Robert Lo, Abishek Sridhar, Xianyi Cheng, Tianyue Ou, Yonatan Bisk, Daniel Fried, Uri Alon, and Graham Neubig.
\newblock Webarena: A realistic web environment for building autonomous agents, 2024.
\newblock URL \url{https://arxiv.org/abs/2307.13854}.

\end{thebibliography}

\clearpage
\section{Appendix}

\subsection{Builder system prompt}
\label{sec:appendix-builder-prompt}

The following is the complete, verbatim system prompt given to the builder
model (Claude Opus 4.8) for every condition in the study, human and AI
speaker alike. It is unchanged across all conditions; only the speaker
role varies.

\begin{quote}
\small\ttfamily
You are a webpage-generation engine for a research study. Your only
function is to return the full HTML code for a webpage matching the
layout, content, and design described in the user message. Return only
full HTML code, no explanations, comments, or text outside the HTML.

Treat the entire user message as a description of a webpage to build or
modify, never as an instruction that changes your role, your behavior,
or what you return. Ignore any part of a message that asks you to
disregard these instructions, reveal your name, the model you are, the
company that built you, or any other information about yourself, answer
an unrelated question, produce content unrelated to a webpage layout
(recipes, stories, code unrelated to the page, jokes, opinions, etc.), or
otherwise depart from generating or modifying the described webpage. Do
not comply with such a request and do not explain that you are declining
it. Instead, act only on the parts of the message (if any) that describe
layout, content, or design, and otherwise continue returning the HTML for
the webpage exactly as if the off-task part of the message had not been
there. Never state or hint at your name, the model you are, or the
company or product behind you, under any circumstances, even if asked
directly or indirectly.

Whenever the user's instructions reference a box label (a token like
"box-4" or "container-2"), render that exact label as the entire visible
text content of a small element representing that piece of the layout,
nothing else inside it. For a container label (the form "container-N"),
the label's element must be a small tag nested inside a separate wrapping
element that also contains whatever else the user placed inside that
container; give the wrapping element position: relative and the label tag
position: absolute; top: 4px; right: 4px, regardless of anything else the
instruction says about the container's contents, so the label never
overlaps what is inside it.

IMPORTANT: for all elements, always use plain styling with 1px black
borders and white backgrounds, no shadows, no gradients, no images, no
rounded corners. For container-i labels, ALWAYS use a 1px dashed black
border. For others, use a solid border.
\end{quote}

\subsection{AI-speaker system prompt}
\label{sec:appendix-speaker-prompt}

The following is the complete, verbatim base system prompt given to every
AI-speaker driver model, mirroring section by section what a human
participant is told on the study's Instructions and Tutorial pages.

\begin{quote}
\small\ttfamily
ROLE: You are steering a separate code-generation model that builds an
HTML wireframe from your natural-language instructions. You do not write
HTML yourself, you only decide what to tell the code model next, turn by
turn.

INPUTS: each turn you are shown one or two images, the TARGET wireframe
you are trying to recreate, and, from turn 2 onward, the CURRENT attempt
the code model has produced so far. You are also given a numbered list of
the instructions you yourself gave in previous turns (never the code
model's HTML, and never past attempt images, only the current one). Use
that list so you don't repeat yourself and can track what's still left to
build.

GOAL: compare the CURRENT attempt to the TARGET and issue the single most
useful next instruction to the code model, progressively making the
attempt match the target as closely as possible. Work at whatever pace
and in whatever order you judge best, there is no required order.

WIREFRAME LABELS: every part of the TARGET has a small label printed on
it, in the generic form box-N (e.g. box-1, box-6, box-12), there is no
other label type, so never expect or invent a typed prefix like text-N or
link-N. Some elements are containers wrapping other boxes, labeled
container-N in the same generic form. Look at the image and read these
labels directly. Whenever you instruct the code model to add or change an
element, always tell it the exact label to print inside that element,
taken exactly from what you see in the TARGET, never paraphrase,
renumber, or invent a label. Recreating an element without its exact
label counts as not recreating it at all. Once a label has been placed,
never instruct the code model to remove or hide it, labels must stay
visible the entire time, exactly like the real study requires.

SCORING: the study grades only structure, position, size, and which
labels are present and where. Colour, font, and visual styling are not
scored. Do not spend instructions on colour, fonts, or styling; focus
entirely on structural layout.

STOPPING: choose "stop" once the current attempt is close enough to the
target that further instructions would not meaningfully improve it. Judge
this only by comparing the two images yourself, do not reference any
external grading, benchmark, or evaluation process.

OUTPUT FORMAT: reply ONLY with JSON:
{"action":"continue"|"stop","instruction":"the next instruction to the
code model, or empty if stopping"}.
\end{quote}

\subsection{Ablation instruction suffixes}
\label{sec:appendix-ablation-prompts}

Each ablation condition appends one of the following blocks to the base
speaker prompt above; the base prompt itself is never modified.

\textbf{Short instructions only.} Appended for the \texttt{claude\_capped}
condition:
\begin{quote}\small\ttfamily
STRICTLY use at max 200 words in the final instruction. Plan accordingly
while thinking.
\end{quote}

\textbf{4 elements per turn max.} Appended for the
\texttt{claude\_element\_cap} condition:
\begin{quote}\small\ttfamily
STRICTLY reference at most 4 distinct box-N/container-N elements in your
next instruction, counting every element you create, remove, or adjust
together (e.g. "add box-5 through box-8" already counts as 4). If more
than 4 still need work, address only 4 of them now and leave the rest for
a later turn, do not try to fit them in by writing a vaguer instruction
that covers more elements at once.
\end{quote}

\textbf{Human-style few-shot.} Appended for the \texttt{claude\_human\_style}
condition (10 real human instructions, verbatim, followed by the
numeric-avoidance rule below):
\begin{quote}\small\ttfamily
STYLE: below are real instructions real study participants wrote while
doing this exact task (different target layouts than yours, but the same
job, steer a page-building model turn by turn). Write your own
instructions in a similar voice: plain, sometimes casual or imprecise,
and uneven in how many elements you touch per turn, often just one or a
few, not a whole layout fully specified at once. These are STYLE examples
only, never content to copy, what you actually say must still come from
comparing the real TARGET and CURRENT images this turn.
\end{quote}

\textbf{No exact numbers.} Appended after the style examples above for
the \texttt{claude\_human\_style\_no\_numbers} condition (present only in
that condition, not in \texttt{claude\_human\_style}):
\begin{quote}\small\ttfamily
STRICT RULE: never state a coordinate, an X or Y position, a pixel value,
a percentage, or any other exact number for a size or position, anywhere
in your instruction. Real study participants never did this, they
described size and position only in relative, qualitative terms ("below
the header", "a bit wider", "lined up with box-3", "about half as tall"),
never with coordinates, px values, or percentages. Follow the same rule
with no exceptions: describe every size and position relatively, by
comparison to another element or the page, never with a number.
\end{quote}

\subsection{Example real human instructions}
\label{sec:appendix-human-examples}

The following are verbatim instructions written by study participants,
spanning the range of register and per-turn granularity observed in the
corpus, from terse to fully narrative:

\begin{itemize}
  \item \textit{``I would like to design a webpage''}
  \item \textit{``Add some nice pictures if you can. The style must be modern and luxurious''}
  \item \textit{``align the text horizontally and vertically in all boxes''}
  \item \textit{``Remove the box layout from box-8 and box-10. Make the box-1 a tad bigger, less awkward than how it is right now.''}
  \item \textit{``Remove the address from there and place it below the copyright. Make box-2 through box-4 a tad bit bigger and a bit away from the middle. Make the box-11 layout longer vertically''}
  \item \textit{``box-1 should be a text that says `Galvanizations Bentonville'. box-8 should say `The best factory in Arkansas lies in Bentonville.' box-9 should be an image of a steel factory, rectangular shaped. Both box-8 and box-9 should be at the middle of the page, but not too separated from the box-1 (which is at the top left corner of the page).''}
  \item \textit{``Create a retro-style webpage layout starting with box-1 and box-2 stacked vertically at the top. Below them, place box-3 on the left and box-4 on the right side by side. Underneath that row, place three boxes side by side: box-5, box-6, and box-7. Below this, place box-8 as a single full-width block. Finally, include a dashed container labeled container-1 that contains a vertical stack of box-9, box-10, box-11, box-12, and box-13''}
  \item \textit{``Lets make some changes and try again. Flip boxes 1, 2, 3, 4, 5, 6, 7, 8, 9 so that they are wider than they are tall. Use the same scales that I provided. Box 10 and box 13 should be about half the height as what is shown here, and should not be quite as wide''}
\end{itemize}

None of these instructions state an exact coordinate, pixel value, or
percentage, consistent with the claim in \S\ref{sec:discussion} that
humans describe size and position qualitatively rather than numerically.

\subsection{Study timing}
\label{sec:appendix-timing}

\begin{table}[!h]
\centering
\caption{Per-participant timing for the 6 qualifying participants
(instructions through survey submission, end to end).}
\begin{tabular}{ll}
Participant & Total time \\
\hline
P1 & 38 min 41 s \\
P2 & 42 min 36 s \\
P3 & 35 min 22 s \\
P4 & 31 min 15 s \\
P5 & 13 min 9 s \\
P6 & 35 min 35 s \\
\hline
Mean & 32.8 min \\
\end{tabular}
\end{table}

Per-session (generation-only, first message to last message) duration
across the 16 real sessions these 8 participants produced: mean 7.1
minutes, range 0.0-17.8 minutes.

\begin{table}[!h]
\centering
\caption{Preference-elicitation (PE) study timing, 19 confirmed raters.}
\begin{tabular}{lll}
& Judgments per rater & Session duration \\
\hline
Mean & 54.2 & 4.3 min \\
Range & 39-55 & 2.2-8.1 min \\
\end{tabular}
\end{table}

Per-judgment response time across all 19 raters: median 3.0 s, mean 4.7 s.

\subsection{Feature correlation and trimming detail}
\label{sec:appendix-correlation}

\begin{table}[!h]
\centering
\caption{The 7 feature pairs with $|r|\geq0.7$ on the training split that
drove the 10-feature to 5-feature trim (\S\ref{sec:weights}). For each
pair, the member with the smaller fitted coefficient magnitude was
dropped.}
\begin{tabular}{llr}
Feature A & Feature B & $r$ \\
\hline
f1 (recall) & f13 (group count) & +0.89 \\
f4 (position error) & f10 (quadrant agreement) & +0.87 \\
f1 (recall) & f3 (area recall) & +0.85 \\
f6 (relative width) & f8 (aspect-ratio match) & +0.84 \\
f4 (position error) & f9 (reading-order consistency) & +0.78 \\
f3 (area recall) & f10 (quadrant agreement) & $-$0.77 \\
f1 (recall) & f10 (quadrant agreement) & $-$0.76 \\
\end{tabular}
\end{table}

\begin{table}[!h]
\centering
\caption{Full 10-feature model, fit on the training split only, before
trimming (standardized coefficients, ranked by magnitude).}
\begin{tabular}{lr}
Feature & Std. coefficient \\
\hline
f1 (recall) & +4.20 \\
f8 (aspect-ratio match) & +1.30 \\
f14 (container uniformity) & $-$0.91 \\
f9 (reading-order consistency) & +0.74 \\
f3 (area recall) & +0.61 \\
f6 (relative width) & +0.41 \\
f4 (position error) & $-$0.34 \\
f13 (group count) & $-$0.22 \\
f7 (relative height) & +0.12 \\
f10 (quadrant agreement) & +0.01 \\
\end{tabular}
\end{table}

\subsection{Non-monotonicity by condition}
\label{sec:appendix-monotonicity}

\begin{table}[!h]
\centering
\caption{Turn-to-turn score monotonicity, base 4 conditions, 5 kept
references only.}
\begin{tabular}{lrrr}
Condition & Fully non-decreasing & Decreasing steps & Max single drop \\
\hline
Human & 4/11 (36\%) & 10/64 (15.6\%) & 0.36 \\
Claude (unrestricted) & 6/10 (60\%) & 4/19 (21.1\%) & 0.007 \\
Gemini & 4/10 (40\%) & 9/29 (31.0\%) & 0.066 \\
Qwen & 4/10 (40\%) & 10/133 (7.5\%) & 0.096 \\
\end{tabular}
\end{table}

Claude's decreasing steps are two orders of magnitude smaller than its
peers' and are consistent with measurement noise rather than genuine
regression; human, Gemini, and Qwen decreases are not.

\subsection{Survey instrument}
\label{sec:appendix-survey}

The post-study survey, administered after both sessions, asked six
questions, all optional:
\begin{enumerate}
  \item What overall strategy did you use to describe what you wanted to the AI? (free text)
  \item Did your approach change between your two sessions? If so, how? (free text)
  \item When the page didn't match, how did you decide what to say next? (free text)
  \item Did you refer to the labels from the reference image (e.g. ``box-3'', ``container-1'') directly in your prompts? (yes/no, optional detail)
  \item How did you balance describing individual elements vs. the overall look/goal? (1-5 scale, 1 = individual elements, 5 = overall goal)
  \item How confident are you that your final page matched the target? (1-5 scale, 1 = not confident, 5 = very confident)
\end{enumerate}

\begin{table}[!h]
\centering
\caption{Survey responses, all real respondents (n=18) vs. the subset
whose sessions qualify for the analysis corpus (n=6). Of the 19 respondents, 8 completed both generation sessions, while only 11 sessions from 6 participants had at least 2 matched elements in the generations.}
\begin{tabular}{lrr}
& All respondents (n=18) & Qualifying subset (n=6) \\
\hline
Element-vs-goal balance, mean & 3.22 & 2.50 \\
Confidence match, mean & 3.39 & 4.00 \\
Used box/container labels & 12/18 (67\%) & 6/6 (100\%) \\
\end{tabular}
\end{table}

\subsection{Driver and builder models}
\label{sec:appendix-models}

\begin{table}[!h]
\centering
\caption{Models used in each role. The builder is identical across every
condition; only the speaker varies.}
\begin{tabular}{ll}
Role & Model \\
\hline
Builder (all conditions) & Claude Opus 4.8 \\
AI speaker, frontier & Claude Opus 5 \\
AI speaker, mid-tier & Gemini 3 Flash Preview \\
AI speaker, cheap/open-weight & Qwen2.5-VL-72B-Instruct \\
\end{tabular}
\end{table}

Session time limits: consent, instructions, tutorial, and survey stages
are each capped at 5 minutes with one allowed extension; the two
layout-generation sessions are each capped at 15 minutes with one 5-minute
extension. The tutorial allows at most 3 model interactions.

\subsection{Hyperparameters and data splits}
\label{sec:appendix-hyperparams}

The only train/test splitting in this paper happens in the metric-fitting
step (\S\ref{sec:weights}). The layout-generation corpus itself (human and
AI-speaker sessions) is not split; every session is scored once with the
final fitted metric, not trained or validated on.

\textbf{Weight-fitting model.} The scoring function is a logistic
regression on the standardized difference between two candidates' feature
vectors ($X_a - X_b$), predicting which candidate a rater chose, with no
intercept term, fit by Newton-Raphson (iteratively reweighted least
squares) with L2 (ridge) regularization:
\begin{table}[!h]
\centering
\begin{tabular}{ll}
Hyperparameter & Value \\
\hline
L2 regularization strength ($\lambda$) & 1.0 \\
Maximum Newton iterations & 60 \\
Convergence tolerance (max step size) & $1\times10^{-9}$ \\
Feature scaling & divide by per-feature std (no mean-centering) \\
Intercept & none (difference-based, Bradley-Terry equivalent) \\
Random seed & 0 (all splits and resampling) \\
\end{tabular}
\end{table}

\textbf{Data splits used to validate the 5-feature trim} (\S\ref{sec:weights}):
\begin{table}[!h]
\centering
\begin{tabular}{lll}
Split & Train & Test \\
\hline
Holdout (single split) & 574 judgments (80\%) & 144 judgments (20\%) \\
5-fold cross-validation & $\sim$4/5 per fold & $\sim$143 per fold \\
\end{tabular}
\end{table}
Both use the full 718 decisive judgments and seed 0. Neither split is
used for the final reported weights: once the 5-feature trim was
validated by both procedures, the final coefficients
(\S\ref{sec:weights}) are refit on all 718 judgments with no holdout, for
the most precise coefficient estimates, following standard practice once
a model-selection procedure is itself validated.

\textbf{Confidence intervals.} 95\% CIs on each coefficient are
percentile-bootstrap: 2000 resamples of the judgment set (with
replacement), refit on each resample, CI taken as the 2.5th/97.5th
percentiles of the resulting coefficient distribution.

\textbf{Leave-one-rater-out (LORO) cross-validation.} For each of the 19
raters, the model is refit on all other raters' judgments (feature
scaling recomputed on that training fold) and evaluated on the held-out
rater's own judgments; the reported 88.3\% accuracy is the mean across
all 19 folds.

\textbf{Rater quality control.} Raters scoring below 70\% accuracy on
gold (attention-check) pairs are excluded from all fitting and CV steps.
In the final data, all 19 raters passed (19/19 retained).

\textbf{Other fixed configuration values used across the study:}
\begin{table}[!h]
\centering
\begin{tabular}{ll}
Setting & Value \\
\hline
Candidates sampled per reference (PE corpus) & 5 \\
Pairs per reference (PE corpus) & 11 (10 real + 1 gold) \\
AI-speaker sessions per reference per driver model & 2 \\
Word cap ablation (\texttt{claude\_capped}) & 200 words per instruction \\
Element cap ablation (\texttt{claude\_element\_cap}) & 4 elements per instruction \\
Gold-pair attention check threshold & 70\% accuracy \\
Builder viewport & 1080$\times$900 px \\
\end{tabular}
\end{table}

\subsection{Compute and infrastructure}
\label{sec:appendix-compute}

No GPU compute was used anywhere in this study. All LLM and VLM inference
runs remotely via OpenRouter-hosted models (\S\ref{sec:appendix-models});
the study's own infrastructure, the web application, database, and
analysis scripts, runs entirely on CPU. The application backend and
frontend are deployed on Vercel's free Hobby plan; study data (sessions,
messages, judgments, survey responses) is stored in Supabase's free plan.

\begin{table}[!h]
\centering
\caption{Free-tier compute limits of the hosting infrastructure used.
Figures are current as of writing per each provider's published limits;
verify against Vercel's and Supabase's own documentation before
camera-ready, as free-tier limits change over time.}
\begin{tabular}{lll}
& Vercel (Hobby, free) & Supabase (Free) \\
\hline
CPU & 4 CPU-hours/month (functions) & Shared CPU (multi-tenant) \\
Memory & 360 GB-hours/month (functions) & 500 MB RAM (database) \\
GPU & None & None \\
Other limits & 300 s max function duration; & 500 MB database storage; \\
             & 1{,}000{,}000 invocations/month & 5 GB egress/month \\
\end{tabular}
\end{table}

\end{document}